\documentclass[12pt]{article}

\pdfoutput=1

\usepackage{moreverb}

\usepackage{lineno,hyperref}

\usepackage{bm}
\usepackage{amsmath}
\usepackage{subfigure}
\usepackage{amsbsy}
\usepackage{amstext}
\usepackage{amsopn}
\usepackage{makeidx}
\usepackage{psfrag}
\usepackage{graphicx}
\usepackage[dvips]{epsfig}
\usepackage{rotating}
\usepackage{amssymb}
\usepackage{listings}
\usepackage{tikz}
\usepackage{color}
\usepackage{algpseudocode}
\usepackage{algorithm}
\usepackage[margin=1.0in]{geometry}
\usepackage[font={footnotesize}]{caption}

\begin{document}



\title{\Large A Comparative Study of 2D Numerical Methods with GPU Computing}

\footnotesize\date{\footnotesize \today}

\author{\normalsize Ben~J.~Zimmerman, Jonathan~D.~Regele$^*$, \& Bong~Wie\\
\footnotesize Department of Aerospace Engineering, Iowa State University, Ames, IA 50011, USA\\
\footnotesize * Corresponding author: \texttt{jregele@lanl.gov} \\ }

\maketitle

\begin{abstract}
\footnotesize

Graphics Processing Unit (GPU) computing is becoming an alternate computing platform for 
numerical simulations. However, it is not clear which numerical scheme will provide the 
highest computational efficiency for different types of problems.
To this end,
numerical accuracies and computational work of several numerical methods are compared 
using a GPU computing implementation. The Correction Procedure via Reconstruction (CPR),
Discontinuous Galerkin (DG), Nodal Discontinuous Galerkin (NDG), Spectral Difference (SD), and Finite Volume (FV) 
methods are investigated using various reconstruction orders. Both smooth and discontinuous cases are considered
for two-dimensional simulations. For discontinuous problems, MUSCL schemes are employed with FV, while 
CPR, DG, NDG, and SD use slope limiting. The computation time to reach a set error criteria and total time 
to complete solutions are compared across the methods. 
It is shown that while FV methods can produce solutions 
with low computational times, they produce larger errors than high-order methods for smooth problems at the same order of accuracy. 
For discontinuous problems, the methods 
show good agreement with one another in terms of solution profiles, and the total computational times between
FV, CPR, and SD are comparable.

\end{abstract}

\vspace{-6pt}

\section{Introduction}
\vspace{-2pt}
Numerical simulation of fluids typically requires high resolution and large computational power. In industrial 
settings, high resolution is usually obtained through the computational domain, and not the computational method itself. This is
because low-order methods such as finite volume (FV) are employed in simulations. 
In this paper, a low-order method implies either $1^{st}$ or $2^{nd}$ order spatial reconstruction, while a
high-order method indicates a solution reconstruction of $3^{rd}$ order and higher \cite{ho}.
This differs from compressible methods, where a low-order method is $1^{st}$ order accurate, and high-order 
is $2^{nd}$ or $3^{rd}$ order accurate.
While it is possible for FV methods to achieve higher-order spatial reconstruction, the computational 
cost becomes high in terms of memory access, especially for unstructured grids \cite{zjsv}. The solution reconstruction requires information from neighboring elements, 
and as the order of accuracy is increased, the number of elements required for communication also increases. 
In contrast, high-order methods only require information at element neighbors, regardless of the order of accuracy. This compact 
nature is appealing to parallel processing, especially Graphics Processing Unit (GPU) computing.

While most practical computations are completed in three-dimensions, \newline
two-dimensional problems are still of interest.
They are even more appealing towards GPUs, whose low memory storage makes computing on high-resolution three-dimensional 
problems a issue. In addition, it is not clear how different numerical methods compare with one-another under GPU implementation,
even for two-dimensional problems.
Various researchers have explored GPU Compute Unified Device Architecture (CUDA) with different numerical methods. 
Implementation of the FV method for GPUs has 
been investigated by Castro \textit{et al.} \cite{castro}, where the governing equations were the shallow water 
equations, and Obenschain \cite{oben} for unstructured meshes. The parallelism of FV per element is limited, 
as solutions are reconstructed along element edges before the volume integration step. In contrast, high-order methods 
have multiple solution states within each element, stored at solution points, which increases parallelism 
per element. The most developed high-order methods to date include Discontinuous Galerkin (DG), 
Nodal Discontinuous Galerkin (NDG), Correction Procedure via Reconstruction (CPR), and Spectral Difference (SD).

Discontinuous Galerkin (DG) \cite{bass,bau,cock,cock0,cock00,reed} was the first high-order method 
introduced to hyperbolic equations. There are mutliple approaches to the DG method, depending on how the 
integration points are chosen. Using Gauss-Legendre points for DG implementation
demands computations of surface and volume integrals at each step. This allows for improved accuracy at a 
cost of increased computational work per step. A more efficient implementation of DG was completed by 
Hesthaven and Warburton \cite{Hes}, which moved the integration points to element edges (NDG). For an in depth 
discussion of the implementation of NDG to GPUs, the reader is directed to the paper by Kl{\"o}ckner et. al.
\cite{klock}. The CPR method was developed to improve efficiency 
of other high-order methods \cite{huynh,wangbook,gao,yu}, which includes the DG method. The CPR approach 
allows the equations to be solved in differential form, removing the added surface and volume integration 
computations present in DG. While this increases the computing speed, the method is not as accurate as the 
DG approach \cite{ho}. CPRs application to GPUs was completed by Hoffmann and Zimmerman \cite{hoff1, zimm}, 
where significant speed-ups are observed. The SD method is a finite difference-like formulation \cite{liu06, may, sun}, 
which uses two sets of points, solution and flux points, where the flux derivative is computed across the 
flux points to update the solution states. The SD methods application to GPUs was completed by Zimmerman \cite{zimm2} for 
a three-dimensional system. 
 
The aforementioned references layout efficient algorithms and implementation techniques for the numerical methods discussed, and 
compare the speeds from GPU to Central Processing Unit (CPU) implementations, where significant speed-up results are shown.
There has been a comparative study done by Yu \textit{et al.} \cite{yu2} on high-order methods using a CPU platform. However, there 
has been no performance assessment of different methods using a GPU platform. Furthermore, there has been no performance 
comparison between high-order methods to FV methods on GPUs.
Thus, the intent of this paper is to perform a fair comparison in two-dimensions of numerical methods and determine the relative performance 
between them in terms of total computing speed and accuracy with GPUs.
The developed approach for two-dimensions should be extended to three-dimensions in subsequent work
in 
order to account for the known shift in computational cost from two to three-dimensions.
To this end, the FV, CPR, DG, NDG, and 
SD methods are all implemented using GPU CUDA, in similar manners from the references discussed above. 
Each method is compared 
at the same order of accuracy and same number of degrees of freedom, with the maximum allowable time-step for a given mesh.
The comparison is for 
the two-dimensional Euler system, for both smooth and discontinuous problems. For discontinuous problems, a shock capturing 
approach is required. For the FV method, the MUSCL scheme \cite{muscl1, muscl, harten} is implemented, while the high-order methods 
use a slope limiter \cite{cock} to limit the order of the solution only at discontinuities.
The present study investigates only quadrilateral elements, where the total number of degrees of freedom 
are held constant between the methods. In addition, each method takes a maximum allowable time-step for stability. 
This plays an important factor 
when considering the work to reach a specified final time, since high-order methods are time-step restricted, and this restriction
increases with the order of accuracy of the scheme \cite{cock9, kub}. 

The paper is organized in the following manner. In section 2, each numerical method implemented is 
discussed briefly. Section 3 outlines the implementation with GPU programming. The results are discussed 
in section 4, where error analysis and computational time information are discussed in detail. Finally, section 5 
draws conclusions from the study.

\section{Numerical Methods}
The hyperbolic conservation law is given by,
\begin{align}
 \frac{\partial {\bm q}}{\partial t} + \vec \nabla \cdot \vec {\bm F}({\bm q}) = 0, 
 \label{hyperbolic}
\end{align}
where ${\bm q}$ is the state vector and $\vec \nabla \cdot \vec{\bm F}({\bm q})$ is the divergence of the inviscid flux vector, which takes the following form,
\begin{align}
 \vec \nabla \cdot \vec {\bm F}({\bm q}) = \frac{\partial {\bm f}({\bm q})}{\partial x} + \frac{\partial {\bm g}({\bm q})}{\partial y}.
\end{align}
For the two-dimensional Euler equations,
$\bm q$ is a vector of the conserved variables, 
\begin{align}
\bm q &= \begin{bmatrix} \rho \\ \rho u \\ \rho v \\ e \end{bmatrix},
\label{eq:state}
\end{align}
and ${\bm f}({\bm q})$ and ${\bm g}({\bm q})$ are flux vectors,
\begin{align}
\bm f(\bm q) = \begin{bmatrix} \rho u \\ p + \rho u^2 \\ \rho u v \\ u(e + p) \end{bmatrix}, \quad \quad
\bm g(\bm q) = \begin{bmatrix} \rho v \\ \rho u v \\ \ p + \rho  v^2 \\ v(e + p) \end{bmatrix}.
\label{eq:flux}
\end{align}
In Eqns. (\ref{eq:state}) and (\ref{eq:flux}), $\rho$ is the density, $u$ is the x-direction velocity, $v$ is the 
y-direction velocity, $e$ is the total energy per unit volume, and $p$ is the pressure. To close the system, 
the ideal gas equation of state is used,
\begin{align}
 p = (\gamma -1)(e - \frac{1}{2} \rho (u^2 + v^2)).
\end{align}
The computational
domain is discretized with non-overlapping elements, each with volume $V_m$. Additionally, each element 
must be transformed into a standard element \cite{jac}.
Within each element, a set of solution points are defined, which stores the solution states. 
\subsection{FV Formulation}
In the FV approach, the solution per element takes on an averaged value. The governing 
equations are integrated over the elements volume, $V_m$,
\begin{align}
 \int_{V_m} \left [ \frac{\partial {\bm q}}{\partial t} + \vec \nabla \cdot \vec {\bm F}({\bm q}) \right ] dV = 0.
\label{fv1}
 \end{align}
The solution average, denoted by ${\bm {\bar q}}_m$ is then defined as 
\begin{align}
 {\bm {\bar {q}}}_m = \frac{1}{V_m}\int_{V_m} \bm {\bar q } dV.
\end{align}
The semi-discretization can then be written in the following well known form for two-dimensional quadrilateral elements,
\begin{align}
 \frac{\partial {\bm {\bar q}}_{i,j}}{\partial t} + \frac{1}{\Delta x} \left [ \bm f_{i+1/2,j} - \bm f_{i-1/2,j} \right ]
 + \frac{1}{\Delta y} \left [ \bm g_{i,j+1/2} - \bm g_{i,j-1/2} \right ] = 0.
\end{align}
In the above formulation, $i$ is the index in the x-direction, while $j$ is the index in the y-direction. To obtain the 
flux at an interface (say $\bm f_{i-1/2,j}$, which is the left interface of the element) left and right solutions need to be reconstructed 
at the elements edge first. Once left and right solutions are found at each interface, a Riemann problem is solved to determine the 
flux value at the interface. The averaged solution is then updated via a time-marching scheme.

\subsection{CPR Formulation}
Here, the CPR method is described. For a full derivation, see \cite{wangbook}.
The formulation of 
the CPR method requires the definition of an arbitrary weighting function ${w}$. By multiplying the 
weighting function to Eqn. (1) and integrating over the domain, Eqn. (\ref{hyperbolic}) becomes
\begin{align}
 \int_{V_m} \left [ \frac{\partial {\bm q}}{\partial t} + \vec \nabla \cdot \vec {\bm F}({\bm q}) \right ] {w} dV = 0.
\label{cpr1}
 \end{align}
%
%
%
By applying the Gauss divergence theorem, Eqn. (\ref{cpr1}) is expanded to be
\begin{align}
  \int_{V_m} \frac{\partial {\bm q}}{\partial t}{w} dV + \int_{\partial V_m} {w} \vec {\bm F}({\bm q}) \cdot {\bm n} dS - \int_{V_m} \vec \nabla {w}
  \cdot \vec {\bm F}({\bm q}) dV = 0.
  \label{cpr11}
 \end{align}
Let ${\bm q}_m$ approximate the solution ${\bm q}$ within the element $V_m$. Furthermore, the solution is assumed to belong to the space of 
polynomials of degree $k$ or less (${\bm q}_m \in P^k$). Thus, Eqn. (\ref{cpr11}) must satisfy the following,
 \begin{align}
  \int_{V_m} \frac{\partial {\bm q}_m}{\partial t}{w} dV + \int_{\partial V_m} {w} \vec {\bm F}({\bm q}_m) \cdot {\bm n} dS - \int_{V_m} \vec \nabla {w}
  \cdot \vec {\bm F}({\bm q}_m) dV = 0.
  \label{cpr12}
 \end{align}
There is no requirement enforced on element edges at this point. The normal flux is replaced with a common Riemann flux to enforce 
element coupling,
 \begin{align}
  \int_{V_m} \frac{\partial {\bm q}_m}{\partial t}{w} dV + \int_{\partial V_m} {w} \vec {\bm F}^n_{com}({\bm q}_m,{\bm q}_{m+})dS - \int_{V_m} \vec \nabla {w}
  \cdot \vec {\bm F}({\bm q}_m) dV = 0.
  \label{cpr13}
 \end{align}
In Eqn. (\ref{cpr13}), ${\bm q}_{m+}$ is the solution outside of element $m$. Next, integration by parts is applied again to the last term in
Eqn. (\ref{cpr13}) to yield
 \begin{align}
  \int_{V_m} \frac{\partial {\bm q}_m}{\partial t}{w} dV + \int_{V_m} {w} \vec \nabla \cdot \vec {\bm F}({\bm q}_m) dV
  + \int_{\partial V_m} {w} \left [ {\bm F}^n_{com} - {\bm F}^n({\bm q_m}) \right ] dS= 0.
  \label{weight}
 \end{align}
In the CPR formulation, the last term in Eqn. (\ref{weight}) is viewed as a penalty term, which can be lifted to a volume integral 
by introducing a correction polynomial $\bm {\delta}_m \in P^k$,
\begin{align}
 \int_{V_m} w \bm{\delta}_m dV =  \int_{\partial V_m} {w} \left [ {\bm F}^n_{com} - {\bm F}^n({\bm q_m}) \right ] dS.
\end{align}
The volume integral formulation of Eqn. (\ref{weight}) is obtained,
 \begin{align}
  \int_{V_m} \left [ \frac{\partial {\bm q}_m}{\partial t} + \vec \nabla \cdot \vec {\bm F}({\bm q}_m)  + {\bm \delta}_m \right ] {w} dV= 0.
 \end{align}
If the conservation law is non-linear, then $\vec \nabla \cdot \vec {\bm F}({\bm q}_m)$ does not generally fall into $P^k$. 
To resolve the non-linear situation, the term $\vec \nabla \cdot \vec {\bm F}({\bm q}_m)$ is projected into $P^k$. 
Then, eliminating the weight and volume integral gives the differential formulation,
 \begin{align}
  \frac{\partial {\bm q}_m}{\partial t} +  \Pi \left [ \vec \nabla \cdot \vec {\bm F}({\bm q}_m) \right ]  + {\bm \delta}_m = 0.
  \label{cpr2}
 \end{align}
%
%
The weighted residual formulation is reduced to a differential one.
Each element must store the solution states at a set of points, called solution points. For the CPR method, 
within an element $V_m$, a set of Legendre-Lobatto solution points are defined, as shown in Fig. \ref{DGpoints} (b). 
At each solution point $j$, Eqn. (\ref{cpr2}) must be true,
\begin{align}
 \frac{\partial {\bm q}_{m,j}}{\partial t} +  \Pi_j \left [ \vec \nabla \cdot \vec {\bm F}({\bm q}_{m}) \right ]  + {\bm \delta}_{m,j} = 0.
\end{align}
Now the calculation of both $\Pi_j \left [ \vec \nabla \cdot \vec {\bm F}({\bm q}_{m}) \right ]$ and the correction polynomial, ${\bm \delta}_{m,j}$, must be completed. 
The inviscid flux divergence follows a chain rule approach (see \cite{tan} for analytical flux derivative). For 
${\bm \delta}_{m}$ formulations, see \cite{gao}. In this work, the correction polynomial is computed using Radau polynomials, which casts $\bm \delta$ into
the DG framework \cite{huynh}, improving accuracy.
The definition of Legendre-Lobatto solution points brings a sense of efficiency into the method. 
The solution points occupy edges of elements, thus no interpolation of information to element edges is required, and element coupling becomes
straightforward.

\subsection{DG Formulation}

\begin{figure}[t]
\centering
\mbox{\subfigure[DG element]{ \includegraphics[width=1.4in]{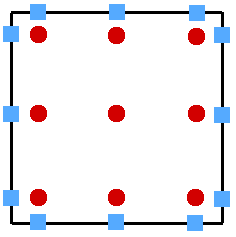}} \quad
      \subfigure[CPR and NDG element]{ \includegraphics[width=1.4in]{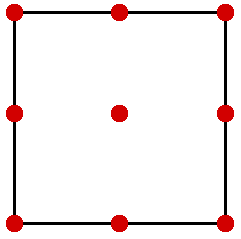}} \quad
      \subfigure[SD element]{ \includegraphics[width=1.36in]{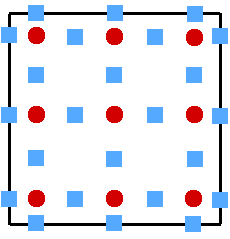} }}
\caption{Solution points (red circles) and flux points (blue squares) locations for $P^2$ solution reconstruction.
(a) Gauss-Legendre solution points with Gauss-Lobatto interface points. Associated with the DG method.
(b) Gauss-Lobatto solution points with coinciding flux points. Associated with CPR and NDG methods.
(c) Gauss-Legendre solution points with Gauss-Lobatto flux points. Associated with SD method.}
\label{DGpoints}
\end{figure}

The DG methods formulation is more straightforward than the CPR method, 
and more information can be found in Ref. \cite{cock}. 
Again, a weighting function $w$ multiplies the conservation law, Eqn. (\ref{hyperbolic}), and is 
integrated over the domain,
\begin{align}
 \int_{V_m} \left [ \frac{\partial {\bm q}}{\partial t} + \vec \nabla \cdot \vec {\bm F}({\bm q}) \right ] {w} dV = 0.
\end{align}
Like the CPR method, integration by parts is performed, and ${\bm q}_m$, which belongs to the space $P^k$, is allowed to approximate the solution on element $V_m$,
\begin{align}
 \int_{V_m} \frac{\partial {\bm q_m}}{\partial t}{w} dV + \int_{\partial V_m} {w} \vec {\bm F}({\bm q_m}) \cdot {\bm n} dS - \int_{V_m} \vec \nabla {w}
 \cdot \vec {\bm F}({\bm q_m}) dV = 0.
 \label{dg1}
\end{align}
The solution and flux polynomials are approximated within each element $i$ over $n$ Gauss-Legendre points as,
\begin{align}
 {\bm q_m} = \sum_{j=1}^{n} {\bm q}_{m,j} \phi_j, \\ \quad
 \vec {\bm F}({\bm q_{m,j}}) = \sum_{j=1}^{n} \vec {\bm F}_{m,j} \phi_j,
 \label{dg2}
\end{align}
where $\phi_j$ are the basis functions. If the basis and weighting functions are equal, then the procedure is Galerkin. 
The surface integral term in Eqn. (\ref{dg1}) couples elements together and the common flux is again calculated via a Riemann solver.
Since the solution points are Gauss-Legendre for the DG method, there is more computational work per time step when 
compared to the CPR method, since solutions must be interpolated to edges before element coupling. Figure \ref{DGpoints} (a) shows 
a typical $P^2$ DG element, where sets of flux points are defined along the edges to communicate solutions.
In addition to the interpolation step, volume and surface integral calculations further increase the computational 
cost of the method.


\subsection{NDG Formulation}

The Nodal DG formulation closely follows the CPR formulation discussed previously. Equation (\ref{hyperbolic}) is 
multiplied by a weighting function and integrated to yield the weak form,
\begin{align}
  \int_{V_m} \frac{\partial {\bm q}_m}{\partial t}{w} dV + \int_{\partial V_m} {w} \vec {\bm F}({\bm q}_m) \cdot {\bm n} dS - \int_{V_m} \vec \nabla {w}
  \cdot \vec {\bm F}({\bm q}_m) dV = 0.
  \label{ndg1}
 \end{align}
A Riemann flux is used to apply element coupling, and replaces $\vec {\bm F}({\bm q}_m) \cdot {\bm n}$ with a common Riemann flux ${\bm F}^n_{com}$,
which uses the current and neighboring element information,
\begin{align}
  \int_{V_m} \frac{\partial {\bm q}_m}{\partial t}{w} dV + \int_{\partial V_m} {w} {\bm F}^n_{com}  dS - \int_{V_m} \vec \nabla {w}
  \cdot \vec {\bm F}({\bm q}_m) dV = 0.
  \label{ndg2}
 \end{align}
For the DG approach, a basis set $w_j$ is chosen for the solution space, where $j$ is the index of each solution point. 
Equation (\ref{ndg2}) is written as the following strong DG form,
\begin{align}
  \frac{\partial}{\partial t}\int_{V_m} w_i {\bm q}_{m,j} w_j dV - \int_{\partial V_m} {w_i} \left [ \vec {\bm F} \cdot \bm{n} - {\bm F}^n_{com}  \right ] dS  +  
  \int_{V_m} {w_i} \vec \nabla \cdot \vec {\bm F}({\bm q}_m) dV = 0.
  \label{ndg3}
 \end{align}
A mass, stiffness, differentiation, and face mass matrices can be formulated, as completed in \cite{klock},
\begin{align}
 M_{i,j} &= \int_{V_m} w_i w_j dV, \\  
 S_{i,j} &= \int_{V_m} w_i \nabla w_j dV, \\
 D_{i,j} &= \left ( M_{i,j} \right )^{-1} S_{i,j}, \\ 
 M^A_{i,j} &= \int_{\partial V_m} w_i w_j dS.
\end{align}
These matrices are used in Eqn. (\ref{ndg3}) to obtain the following formulation,
\begin{align}
 \frac{\partial {\bm q_{m,j}}}{\partial t} + D \left [\vec {\bm F}({\bm q}_m) \right ] - L\left [ \vec {\bm F} \cdot \bm{n} - {\bm F}^n_{com}  \right ]_A = 0.
\end{align}
The matrix $L$, or lifting matrix, acts on the facial degrees of freedom on face $A_m$. It combines the mathematical aspects of applying the mass matrix on the face, 
lifting the facial integral to volume integral, and finally applying the inverse mass matrix. Much like CPR, this method also uses Gauss-Lobatto quadrature as the solution 
points (see Fig. \ref{DGpoints} (b)), simplifying the element communication step. 

\subsection{SD Formulation}
The SD scheme employs a finite-difference like approach on the conservation laws. The solution is assumed to be in the space $P^k$, while the flux is assumed
to be in the space $P^{k+1}$. A set of solution points and flux points are defined within each element. Figure \ref{DGpoints} (c) illustrates the point locations in 
a SD $P^2$ element. Note how an extra flux point is required per direction for the flux polynomial. 
The solution states are stored at the solution points 
while the flux points compute the flux terms. Let $h(\xi)$ define the degree $k$ Lagrange polynomial at the solution points and $l(\xi)$ be the degree $(k+1)$ polynomial 
at the flux points. The coordinates $(x,y)$ are transformed into standard coordinates $(\xi, \eta)$. The solution is reconstructed as tensor products of 
two one-dimensional polynomials,
\begin{align}
 {\bm q} = \sum_{j=1}^{k+1} \sum_{i=1}^{k+1} { \bm q}_{i,j} h_i(\xi) h_j(\eta).
\end{align}
In a similar manner, the reconstructed flux polynomials are formulated as
\begin{align}
 {\bm f} = \sum_{j=1}^{k+1} \sum_{i=0}^{k+1} { \bm f}_{i+1/2,j} l_{i+1/2}(\xi) h_j(\eta), \\
 {\bm g} = \sum_{j=0}^{k+1} \sum_{i=1}^{k+1} { \bm g}_{i,j+1/2} h_i(\xi) l_{j+1/2}(\eta).
 \end{align}
In this formulation, $i$ and $j$ indicate the points in $x$ and $y$ directions respectfully. The flux polynomials are 
only continuous within each element. To resolve the discontinuous interface, a Riemann solver is applied at flux points 
on the interfaces to provide element coupling. Once the fluxes at the interface are augmented to a common value, 
the flux derivatives are evaluated as
\begin{align}
 \frac{\partial  {\bm f}}{\partial \xi} = \sum_{r=0}^{k+1}  {\bm f}_{r+1/2,j} l'_{r+1/2}(\xi_i), \\
 \frac{\partial  {\bm g}}{\partial \eta} = \sum_{r=0}^{k+1}  {\bm f}_{i,r+1/2} l'_{r+1/2}(\eta_j).
\end{align}
The term $l'(\xi_i)$ is the derivative of the flux points lagrange polynomial evaluated at the solution point locations $\xi_i$.
 
\subsection{Shock Capturing}
To resolve solution discontinuities, the low-order and high-order methods follow two approaches. For the FV method, the second and 
third order MUSCL schemes are implemented, which is applied during the reconstruction of the solution at element 
interfaces. The slopes of the reconstructed solutions are limited with the \textit{minmod} limiter \cite{minmod}. 
For second order reconstruction, the second order MUSCL scheme is applied, while the third order MUSCL scheme is 
selected for third order reconstruction.

For the high-order methods, the same technique is applied for all schemes, which uses a \textit{minmod} 
limiter (similar to FV) to find troubled elements and apply slope limiting. 
The updated solution is interpolated (if need be) to element edges. Once interpolation is completed, the \textit{minmod} limiter 
is applied to reconstruct a second solution based on cell averaged values to the edge. If the difference in these two values 
is greater than a certain threshold (numerical experiments indicate $> 1.0 \times 10^{-3}$ gives good solutions) then the cell is marked 
for limiting, where the new solution is, 
\begin{align}
 {\bm q}_{m,j} = \bar {\bm q}_{m} + (x_{m,j} - x {\rm 0}) {\rm minmod}\left (\frac{\bar {\bm q}_{m+1} - \bar {\bm q}_{m}}{h}, \frac{\bar {\bm q}_{m} - \bar {\bm q}_{m-1}}{h} \right ).
 \label{limu} 
 \end{align}
In Eqn. (\ref{limu}), ${\bm q}_{m,j}$ is the solution in element $m$ and solution point $j$, $x_{m,j}$ is the location of the solution point, $x {\rm 0}$ is 
the element midpoint, $h$ is the element size, and $\bar {\bm q}_m$ is the averaged solution in an element. This scheme results in a second order 
reconstruction which can be applied to any of the high-order methods discussed in this paper.

\section{GPU CUDA Overview and Implementation}

Before discussing the implementation of methods into GPU CUDA, a brief overview of GPU computing is presented, to 
give the reader a basic understanding of some conventions and algorithms on the GPU. For a more complete 
discussion, refer to the NVIDIA CUDA programming guide \cite{nvidia}. 

\subsection{CUDA Overview}

Graphics computing is aimed toward image rendering, a largely parallel task. GPUs are built around 
streaming multiprocessors to complete tasks, which execute hundreds of independent threads. The 
multiprocessors launch blocks, containing threads, running in parallel. Threads within a block are 
allowed to share information through the GPU's shared memory (each GPU has a limit on the amount of 
shared memory available). This architecture is coined as single-instruction-multiple-thread (SIMT) 
architecture.
The blocks are executed through a grid, where no communication is allowed 
between the threads, and there is no guarantee of which block will finish first. Only after every blocks
work is completed can the grid be viewed, and data can be analyzed or seen by other threads if 
the appropriate memory was written into a GPU's global memory. Global memory can be seen by all threads in all blocks
on the GPU, and every thread can write to this memory. However, the cost to write to this memory location can be 
high (hundreds of clock cycles). So writes into this memory should be completed only when necessary. Global memory 
can be bound to texture memory to hasten read access. In this implementation, all global memory is also allotted 
space in the texture memory. Finally, shared memory is used when threads in a block need to communicate information 
to one another. Typical usage of this occurs during for loops, where one thread needs the information of other 
threads to perform computations, such as derivatives.

There are a few rules to follow when writing CUDA code to help optimize computing speed. 
\begin{itemize}
 \item The usage of shared memory should be minimized and reused when possible.
 \item The storage locations of memory should compliment the SIMT architecture.
 \item Threads should be synchronized rarely and in optimal locations.
 \item Each thread should write to global memory only once.
\end{itemize}
Some are quite obvious, such as the recycling of shared memory and location of barriers. For storage 
order of memory, consider the following case: Let's use the Euler system and assume a memory storage 
where at a single point, memory position 0 is conservation of mass, memory position 1 and 2 are conservation of 
momentum in x and y, and memory position 3 is conservation of energy. Now, let thread 0 read memory position 0, thread 
1 read memory position 1, and so on. One can observe that evaluating components of the field is completed by evaluating 
different expressions, which means different code, inefficient for SIMT architecture. A better solution is to let thread 0 
access memory position 0 of the point, and thread 1 access the memory position 0 of another point, which allows the same 
expression to be computed by the threads. The final item, one global write, is also self explanatory, since each access to global 
memory is expensive. It is noted, however, that in some cases this cannot be followed, and allowing multiple writes to global
is cheaper than splitting the algorithm.

Some conventions are now listed to simplify the algorithms presented, and assist the reader. Threads and blocks are allowed 
to be multi-dimensional, and have the indexes $t_x$, $t_y$, $t_z$, $b_x$, and $b_y$ (threads can have three indexes while blocks
can have up to two). Memory locations are presented in the following manner: Assume some code variable $u$, which can be in any of the 
following memory locations based on the superscript. 
If the variable has no superscript, it resides in the local memory to the 
thread. The other three locations are denoted by superscripts $g$, $t$, and $s$ to represent global, texture, and shared memory
space respectfully.
In addition, any memory reads or writes with indexes will be denoted in the following manner: If stored   
memory needs to be read (say $c$ is the pointer or array which holds the information), and the indexes depend on $i, j,$ and $k$, then let $v = c[(i,j,k)]$. 
Meaning, $v$ now reads information in $c$ at 
an index location which depends on $i, j$, and $k$ (not a three dimensional array or pointer).

\subsection{CUDA Implementation}

\begin{figure}[t]
\centering
\includegraphics[width=4.5in]{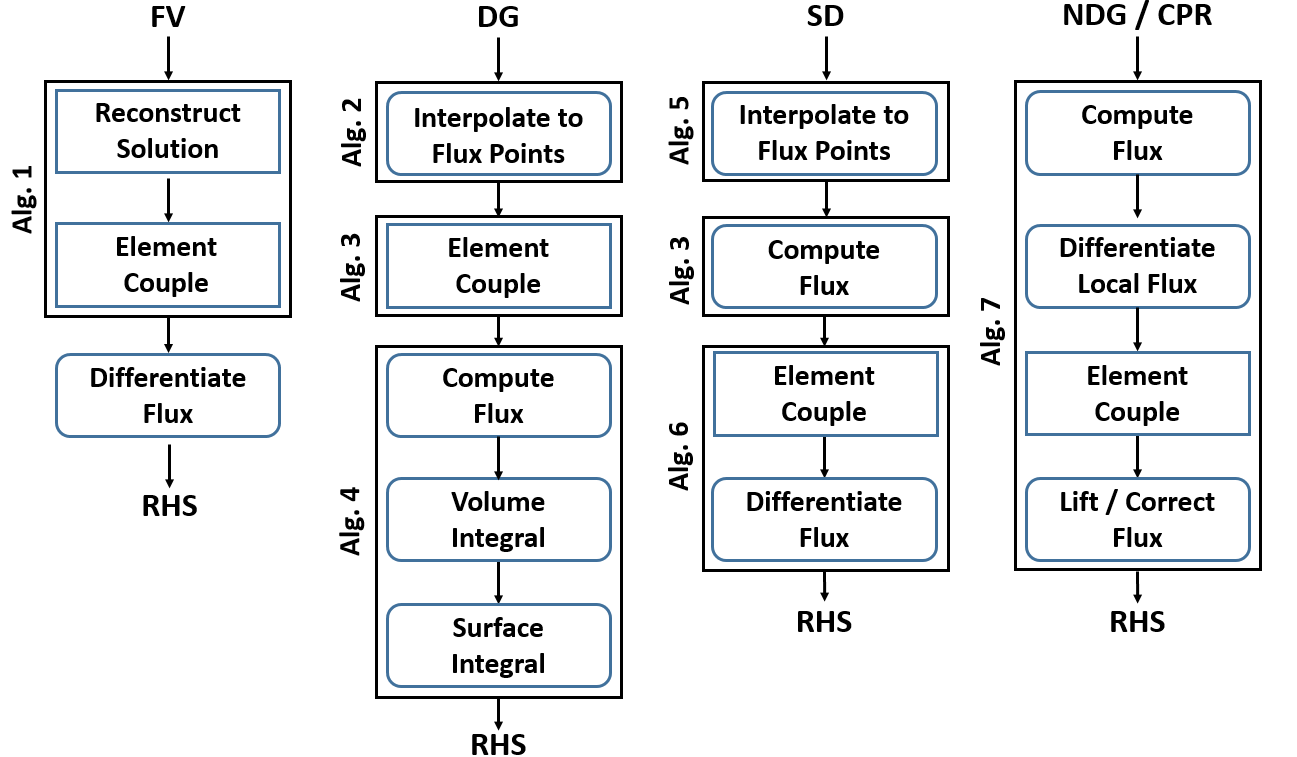}
\caption{Overview of methods. Rectangles indicate non-local operations, while other shapes indicate local operations.}
\label{steps}
\end{figure}

Now that the basic idea is presented, the GPU implementation of each method is presented. Each
methods entire implementation will not be discussed, only the residual update and shock capturing algorithm. 
The remaining functions were 
implemented according to the algorithms found in Ref. \cite{zimmThesis}.
An overview of each method's steps are outlined in Fig. \ref{steps}.
For each method, the local and non-local operations are shown. A local operation means all information to complete 
the operation is contained within the element, while non-local means communication must occur between elements. Note that
the number of operations listed does nor correlate with the number of functions required. Some operations, local and non-local, can 
be combined into one function to reduce memory loading and multiple sweeps through the domain.

\subsection{FV CUDA}
The FV method can be separated into two seperate kernels to update the residual. As shown in Fig. \ref{steps}, one kernel 
reconstructs the solution and provides element coupling (both non-local operations) whose output feeds into the flux differentiation kernel.
\algnewcommand{\LineComment}[1]{\State \(\triangleright\)  #1}
\begin{algorithm}[t]\footnotesize
\caption{FV\_Reconstruct}
\begin{algorithmic}
\LineComment{Faces in element}
\State $t_x$    = threadIdx.x
\LineComment{Current global face}
\State $k = \text{blockIdx}.x * \text{blockDim}.x + i$
\If{$k < \text{n}_{e}$}
\LineComment{Gather information from neighbors}
\State $q_{e_1}[(0...n_v)] = q^t[id_{e_1}(t_x,k)]$ 
\State $q_{e_2}[(0...n_v)] = q^t[id_{e_2}(t_x,k)]$ 
\State ...
\LineComment{Reconstruct left and right solutions}
\State $q_L[(0...n_v)] = f(q_{e_1}, q_{e_2} ...)$
\State $q_R[(0...n_v)] = f(q_{e_1}, q_{e_2} ...)$
\LineComment{Compute the interface flux}
\State $\text{InterfaceFlux}(q_L, q_R, f_n)$
\LineComment{Store interface flux into global memory}
\State $f^g[k + (0...n_v)*n_f] = f_n[(0...n_v)]$
\EndIf
\end{algorithmic}
\end{algorithm}
The \textit{FV\_Reconstruct} algorithm is outlined in Algorithm 1, which reconstructs the left and right solutions at faces and computes 
the Riemann flux at the face. Our implementation uses strictly texture memory and registers, with one write to global memory to finish the 
algorithm. The threads are defined as faces, $t_x$, in the domain. In all algorithms, multiple elements (or faces) are 
calculated in one block, increasing the parallelism of the algorithm. The variables $n_v$ and $n_e$ denote the number of state variables 
and number of elements respectfully. Depending on the degree of the reconstruction polynomial, an appropriate amount of information from 
neighbors is loaded (the index $e_1$ denotes element 1). Once the data is loaded, the appropriate reconstruction formula is applied and the 
flux at the interface is computed and stored.

To compute the flux derivative, multiple elements are computed per block, and each thread reads the appropriate flux information from 
texture memory to compute the flux derivative in the element (computed from \textit{FV\_Reconstruct}). The FV method, while simplistic, demands memory transfers from neighboring elements 
in the domain, which is the major bottleneck in the method.

\subsection{High-Order Methods CUDA}
For GPU implementation of each high-order method, a general algorithm is presented to compute 
the flux derivative. The CPR and NDG methods will be lumped together, as implementation of the two is quite similar. 
The variables $n_{sp1d}$, $n_{fp1d}$,  $n_{sp}$, $n_{fp}$, and $n_{ep}$ denote the number of 
solution points in one dimension, number of flux points in one dimension, total number of solution points in an element, total number of flux points in an element, 
and total number of edge points respectfully.

\subsubsection{DG CUDA}

As shown in Fig. \ref{steps}, the decomposition of the DG residual update requires three kernels. The three algorithms (corresponding 
to three kernels) interpolate the information from solution points to flux points on element edges, couple the elements via a Riemann flux, 
and compute the volume and surface integrals using information stored at both solution and flux points.
\begin{algorithm}[H]\footnotesize
\caption{DG\_Interpolation}
\begin{algorithmic}
\LineComment{Point on element face}
\State $t_x$    = threadIdx.x
\LineComment{Current face in block}
\State $t_y$ = threadIdx.y
\LineComment{The global face}
\State $f = \text{blockIdx}.x * \text{blockDim}.y + t_y$
\If{$f < \text{n}_{f}$}
\LineComment{Gather index information on face}
\State $m = id^t_f(t_x,f)$
\For{$l=0$ to $\text{n}_{sp1d}$}
\LineComment{Read solution point information and operate}
\State $id = id^t_{sp}[(m,l)]$
\State $q_l[(0...n_v)] = q_l[(0...n_v)] + cint^t[l] * q^t[id]$
\EndFor
\State $q^g_l[(t_x,f,0...n_v)] = q_l[(0...n_v)]$
\EndIf
\end{algorithmic}
\end{algorithm}

\begin{algorithm}[H]\footnotesize
\caption{DG\_Couple}
\begin{algorithmic}
\LineComment{Point on element face}
\State $t_x$    = threadIdx.x
\LineComment{Current face in block}
\State $t_y$ = threadIdx.y
\LineComment{The global face}
\State $f = \text{blockIdx}.x * \text{blockDim}.y + t_y$
\If{$f < \text{n}_{f}$}
\LineComment{Read data (normals, cell indexes)}
\State ...
\LineComment{Read in left and right solution)}
\State $q_{L}[(0...n_v)] = q^t[(id_{e_1}(t_x,f),0...n_v)]$ 
\State $q_{R}[(0...n_v)] = q^t[(id_{e_2}(t_x,f),0...n_v)]$ 
\LineComment{Compute the interface flux}
\State $\text{InterfaceFlux}(q_L, q_R, f_n)$
\LineComment{Store interface flux into global memory}
\State $f^g_n[(t_x,f,0...n_v)] = f_n[(0...n_v)]$
\EndIf
\end{algorithmic}
\end{algorithm}

\begin{algorithm}[H]\footnotesize
\caption{DG\_Flux}
\begin{algorithmic}
\LineComment{Solution points, current element in block, global block}
\State $t_x$    = threadIdx.x
\State $t_y$ = threadIdx.y
\State $k = \text{blockIdx}.x * \text{blockDim}.y + t_y$
\If{$k < \text{n}_{e}$}
\LineComment{Read state at solution points}
\State $q[(0...n_v)] = q^t[(t_x,k,0...n_v)]$
\LineComment{Compute flux into shared memory}
\State $f^s[t_x,t_y,(0...nv)] = F(q[(0...n_v)])$
\State $g^s[t_x,t_y,(0...nv)] = G(q[(0...n_v)])$
\LineComment{Threads need to wait for shared memory to fill}
\State syncthreads()
\LineComment{Compute volume integral using shared memory}
\For{$l=0$ to $\text{n}_{sp}$}
\LineComment{Stiffness matrix coefficients}
\State $(S_x, S_y) = (S^t_x, S^t_y)[(t_x,l)]$
\State $Vol[(0...n_v)] = Vol[(0...n_v)] + V([S_x, S_y, f^s, f^y])$
\EndFor
\LineComment{Surface integral next}
\For{$l=0$ to $\text{n}_{fp}$}
\LineComment{Read integration term}
\State $I = I^t[(t_x,l)]$
\LineComment{Read in flux at interface points and compute surface integral}
\State $f_n[(0...n_v)] = f^t_n[(l,k,0...n_v)]$ 
\State $Sur[(0...n_v)] = Sur[(0...n_v)] - f_n[(0...n_v)]*I$
\EndFor
\LineComment{Assemble flux derivative and store}
\State $Res^g[(t_x,k,0...n_v)] = M^{-1}*(Vol[(0...n_v)] + Sur[(0...n_v)])$
\EndIf
\end{algorithmic}
\end{algorithm}

The \textit{DG\_Interpolation} kernel runs threads along each point in all the faces in the domain. At each face, 
the solution point information is read from texture memory, which serves as an index to read the required state
at the solution points. The solution states and interpolation coefficients ($cint$) are read from textured memory to 
perform the indicated operation, which is stored in global memory for future access. Note that $cint$ is used in other algorithms to indicate interpolation coefficients,
but the coefficients are not the same between the algorithms. 
To couple the elements, an 
even simpler kernel (\textit{DG\_Couple}), only demands the left and right information at interfaces, obtained in Algorithm 2.
Boundary conditions are imposed on $q_R$ if necessary.

In Algorithm 4, the threads run on solution points within elements, and multiple elements 
are packed within a thread block. A sufficient amount of shared memory is allocated for storage of the flux, and 
threads are halted while the memory is loaded. Shared memory in this case offers high computational efficiency, since the 
volume integration loop requires information at other solution points in the element. The surface integral is computed 
in a similar manner, without the use of shared memory. The flux derivative is assembled and stored in the GPUs global memory.

\subsubsection{SD GPU}

Like DG, SD also decomposes nicely into three seperate kernels as shown in Fig. \ref{steps}: Interpolations, coupling, and flux computation. Two major 
differences in implementation are the following: SD has no volume integration and each element has interior flux points (not 
just on the edges). This aspect makes the interpolation more expensive in terms of operations and storage, but the final flux 
evaluation cheaper.

\begin{algorithm}[H]\footnotesize
\caption{SD\_Interpolation}
\begin{algorithmic}
\LineComment{Flux points in one direction}
\State $t_x$    = threadIdx.x
\LineComment{Current element in block}
\State $t_y$ = threadIdx.y
\LineComment{The global element}
\State $k = \text{blockIdx}.x * \text{blockDim}.y + t_y$
\If{$f < \text{n}_{e}$}
\LineComment{Read states into shared memory}
\If{$j<\text{n}_{sp}$}
\State $q^s[(t_x,t_y,0...n_v)] = q^t[t_x,k,0...n_v)]$
\EndIf
\State syncthreads()
\LineComment{Build polynomial at flux points (x-direction)}
\For{$l=0$ to $\text{n}_{sp1d}$}
\State $q_x[(0...n_v)] = q_x[(0...n_v)] + cint^t[(l,t_x)] * q^s[(l,t_x,t_y,0...n_v)]$
\EndFor
\LineComment{Compute flux terms}
\State $f_x[(0...nv)] = F(q_x[(0...n_v)])$
\LineComment{Store states and flux at flux points}
\State $q_{x,y}^g[(t_x,k,0...n_v)] = q_x[(0...n_v)]$
\State $f_{x,y}^g[(t_x,k,0...n_v)] = f_x[(0...n_v)]$
\LineComment{Repeat for y-direction}
\State ...
\EndIf
\end{algorithmic}
\end{algorithm}

The interpolation must be completed in each coordinate direction separately, and both the solution states and flux terms 
must be stored in global memory for future use. The \textit{SD\_Interpolation} kernel, outlined in Algorithm 5, 
sets each thread as a flux point in an element, and takes 
time to load the solution states into shared memory. Note that for SD, $n_{sp} < n_{fp}$, regardless of order of accuracy.
This shared memory will be used for both interpolation in the x and y-directions. Once storage for x-coordinates are completed,
the y-direction terms are computed. For the coupling of elements, the reader is referred back to Algorithm 3. Each element 
now has left and right solutions available in global memory access, and the DG algorithm can be used to store the interface flux. The 
only difference is the location of that stored flux. Rather than $f_n^g$, it is simply stored in $f_{x,y}^g$, overwriting the original 
memory from the kernel \textit{SD\_Interpolation}. 

\begin{algorithm}[H]\footnotesize
\caption{SD\_Flux}
\begin{algorithmic}
\LineComment{Flux points in one direction $(n_{sp1d}*n_{fp1d})$}
\State $t_x$    = threadIdx.x
\LineComment{Current element in the block and global element}
\State $t_y$ = threadIdx.y
\State $k = \text{blockIdx}.x * \text{blockDim}.y + t_y$
\LineComment{Solution point and flux point indexes}
\State $isp = mod(t_x,n_{sp1d})$
\State $ifp = t_x/n_{sp1d}$
\If{$k < \text{n}_{e}$}
\LineComment{Read fluxes into shared memory}
\State $id_x = id_x^t(isp, ifp)$
\State $id_y = id_y^t(isp, ifp)$
\State $f_x^s[(id_x,t_y,0...n_v)] = f^t_{x,y}[(id_x,k,0...n_v)]$
\State $f_y^s[(id_y,t_y,0...n_v)] = f^t_{x,y}[(id_y,k,0...n_v)]$
\State syncthreads()
\LineComment{Now only run on solution points}
\If{$ifp < n_{sp1d}$}
\LineComment{Flux differentiation on solution points}
\For{$l=0$ to $\text{n}_{fp1d}$}
\LineComment{Derivative coefficients}
\State $c_x = c_x[(isp, l)]$
\State $c_y = c_y[(ifp, l)]$
\LineComment{Flux derivative per direction}
\State $dF_x[(0...n_v)] = dF_x[(0...n_v)] + c_x*f_x^s[(id_x,l,t_y,0...n_v)]$
\State $dF_y[(0...n_v)] = dF_y[(0...n_v)] + c_y*f_y^s[(id_y,l,t_y,0...n_v)]$
\EndFor
\State $Res^g[(t_x,k,0...n_v)] = dF_x[(0...n_v)] + dF_y[(0...n_v)]$
\EndIf
\EndIf
\end{algorithmic}
\end{algorithm}

The kernel \textit{SD\_Flux} (Algorithm 6) gathers all the flux terms and computes the derivative at the solution points. The threads are allowed to run across solution and 
flux points in one direction. This allows the flux to be loaded into shared memory in the two coordinate directions $(x,y)$. The extra flux point
thread is stopped, and the algorithm continues to run only on solution points. Implementations of the method were performed without shared memory, and threads only 
operated across solution points reading the flux values from textured memory. The presented algorithm was found to be around $1\%$ faster.
The derivative of the flux is completed across the flux points and stored at the solution points (these coefficients are all in $c_x$ and $c_y$). 
The final results are written to global memory for time-stepping.

\subsubsection{CPR / NDG CUDA}

Both CPR and NDG methods have the unique property that solution and flux point coincide
with one another, which enables the entire algorithm to be written in one GPU kernel, as outlined in Fig. \ref{steps}. The major differences between the two 
methods are illustrated within the algorithm presented. For CPR, the solution states are loaded into memory, the \textbf{for} loop computes the 
solution derivatives, and the projections are computed and stored. NDG requires the flux values to be stored and the flux derivatives computed 
within the \textbf{for} loop.

\begin{algorithm}[H]\footnotesize
\caption{CPR\_Flux / (NDG\_Flux) Part 1}
\begin{algorithmic}
\LineComment{Max of flux points or solution points}
\State $t_x$    = threadIdx.x
\LineComment{Current element in the block and global element}
\State $t_y$ = threadIdx.y
\State $k = \text{blockIdx}.x * \text{blockDim}.y + t_y$
\LineComment{Solution points in x and y directions}
\State $i_x = mod(t_x,n_{sp1d})$
\State $i_y = t_y/n_{sp1d}$
\If{$k < \text{n}_{e}$}
\LineComment{Operate on solution points first}
\If{$t_x < n_{sp}$}
\LineComment{Only CPR - Load solution into shared memory from texture}
\State $q^s[(t_x,t_y,0...n_v)] = q^t[(t_x,k,0...n_v)]$
\LineComment{Only NDG - Load solution from texture and store flux values}
\State $f^s[(t_x,t_y,0...n_v)] = f(q^t[(t_x,k,0...n_v)])$
\State syncthreads()
\For{$l=0$ to $n_{sp1d}$}
\LineComment{Only CPR}
\State $c_x = c_x^t[(l,i_x)]$
\State $c_y = c_y^t[(l,i_y)]$
\State $dq_x[(0...n_v)] = dq_x[(0...n_v)] + c_x*q^s([i_x,l,t_y,0...n_v)]$
\State $dq_y[(0...n_v)] = dq_y[(0...n_v)] + c_y*q^s([i_y,l,t_y,0...n_v)]$
\LineComment{Only NDG}
\State $d_x = d_x^t[(l,i_x)]$
\State $d_y = d_y^t[(l,i_y)]$
\State $df_x[(0...n_v)] = df_x[(0...n_v)] + d_x*f^s([i_x,l,t_y,0...n_v)]$
\State $df_y[(0...n_v)] = df_y[(0...n_v)] + d_y*f^s([i_y,l,t_y,0...n_v)]$
\EndFor
\LineComment{Only CPR - Compute projections}
\State $Proj[(0...n_v)] = P(dq_x, dq_y)$
\EndIf
\State syncthreads()
\State ...
\algstore{bkbreak}
\end{algorithmic}
\end{algorithm}

The algorithm sets threads to operate over flux points or solution points, \newline 
whichever is larger (the algorithm can then switch to operating 
on the other set within the kernel). First, operations are completed over solution points, reading in the solution states and storing 
the states (flux values for NDG) into shared memory. The shared memory is used in computing derivatives of the states in CPR, or the flux for 
NDG. The chain rule is used for the flux derivative in CPR. 

\begin{algorithm}[H]\footnotesize
\caption{CPR\_Flux / (NDG\_Flux) Part 2}
\begin{algorithmic}
\algrestore{bkbreak}
\State ...
\If{$t_x < n_{fp}$}
\LineComment{Couple elements (see DG\_Couple)}
\State ...
\LineComment{Store normal flux difference into shared memory ($f^s$ for NDG)}
\State $q^s[(t_x,t_y,0...n_v)] = f_x[(0...n_v)]*n_x + f_y[(0...n_v)]*n_y - f^n[(0...n_v)]$
\EndIf
\State synctheads()
\If{$t_x < n_{sp}$}
\LineComment{Get number of updates per solution point}
\State $n_{upd} = n^t_{upd}[t_x]$
\LineComment{Correct the normal flux (Lift the flux for NDG)}
\For{$l=0$ to $n_{upd}$}
\LineComment{Locations for correction (lifting)}
\State $id = id^t[(t_x,l,k])$
\LineComment{Only CPR}
\State $Corr[(0...n_v)] = Corr[(0...n_v)] - c^t[id] * q^s[(t_x,t_y,0...n_v)]$
\LineComment{Only NDG}
\State $Lift[(0...n_v)] = Lift[(0...n_v)] - L^t[id] * f^s[(t_x,t_y,0...n_v)]$
\EndFor
\LineComment{Only CPR}
\State $Res^g[(t_x,k,0...n_v)] = Proj[(0...n_v)] + Corr[(0...n_v)]$
\LineComment{Only NDG}
\State $Res^g[(t_x,k,0...n_v)] = df_x[(0...n_v)] + df_y[(0...n_v)] + Lift[(0...n_v)]$
\EndIf
\EndIf
\end{algorithmic}
\end{algorithm}

The second part of the algorithm switches the threads to operate on flux points. The coupling of 
the elements is straightforward, as information is already on element interfaces. The normal 
flux difference on the flux points is stored into the same shared memory space from Algorithm 7. 
The threads are switched a final time to operate on solution points, where, based on the method, the normal 
flux difference stored in shared memory is corrected or lifted and used to update the residual.

\subsubsection{Shock Capturing}

The shock capturing algorithm for FV differs significantly from the other methods. In Algorithm 1, the left and 
right states at element interfaces is computed. Immediately following this step, the solutions can be limited 
using an appropriate slope limiting routine following second or third order MUSCL reconstruction.

Unlike the FV method, the approach used in high-order methods requires
extra sweeps through the computational domain. The slope limiting requires 
solution averages at elements, hence the high-order methods must first build the averaged solution within each element before 
any slope limiting can be applied.

\begin{algorithm}[H]\footnotesize
\caption{Average}
\begin{algorithmic}
\LineComment{The current solution state}
\State $t_x$    = threadIdx.x
\LineComment{The current element in the block}
\State $t_y$    = threadIdx.y
\LineComment{Current global element}
\State $k = \text{blockIdx}.x * \text{blockDim}.y + t_y$
\If{$k < n_e$}
\For{$l=0$ to $n_{sp}$}
\LineComment{Build average}
\State $q_m[t_x] = c_m^t[l] * q^t[(t_x,l,k)]$
\EndFor
\State $q_m^g[(t_x,k)] = q_m[t_x]$
\EndIf
\end{algorithmic}
\end{algorithm}

The \textit{Average} kernel (Algorithm 9) builds the solution averages using information from texture memory and stores the result in global memory space.
These averages are used in \textit{Limit} kernels (Algorithms 10 and 11) for limiting.

\begin{algorithm}[H]\footnotesize
\caption{Limit Part 1}
\begin{algorithmic}
\LineComment{Solution points}
\State $t_x$    = threadIdx.x
\LineComment{The current element in the block}
\State $t_y$    = threadIdx.y
\LineComment{Current global element}
\State $k = \text{blockIdx}.x * \text{blockDim}.y + t_y$
\If{$k < n_e$}
\LineComment{Run through points on edges}
\If{$j < n_{ep}$}
\LineComment{Interpolate to edge if necessary}
\State $q_l = $ ...
\LineComment{Load index locations of neighboring elements}
\State $(i_1,i_2) = $...
\LineComment{Construct the minmod at the edge}
\State $q_e = q_m^t[k] + minmod(q_m^t[k]-q_l,q_m^t[i_2]-q_m^t[k],q_m^t[k]-q_m^t[i_1])$ 
\LineComment{Check if solution is large}
\If{$|q_l - q_e| > \epsilon$}
\State $mark=1$
\EndIf
\LineComment{Store the mark into shared memory at edge}
\State $tmp_s[(t_x,t_y)] = mark$
\State syncthreads()
\State ...
\algstore{bkbreak}
\end{algorithmic}
\end{algorithm}

The \textit{Limit Part 1} kernel starts by 
switching the threads to run on the edge points of elements. At each edge point, the solution is read 
from memory (CPR or NDG) or is interpolated from solution point information (DG or SD). Then 
the minmod limiter is applied at the edge to detect if the element has a discontinuity, and if there is one, the 

\begin{algorithm}[H]\footnotesize
\caption{Limit Part 2}
\begin{algorithmic}
\algrestore{bkbreak}
\State ...
\State $mark = 0$
\LineComment{Each edge point runs through the others edge points}
\For{$l=0$ to $n_{ep}$}
\State $mark = mark + tmp_s[(l,t_y)]$
\EndFor
\State $mark_s[t_y] = mark$
\EndIf
\State syncthreads()
\If{$j < n_{sp}$}
\If{$mark_s[t_y] > 0$}
\LineComment{Apply slope limiting now}
\State ...
\EndIf
\EndIf
\EndIf
\end{algorithmic}
\end{algorithm}

Every edge point needs to see the markings of the others, which is completed using a summation. This way, an element 
with at least one troubled point will give each edge point a value of one. The marking is sent into shared memory so 
the information can be communicated when the threads switch to operate across the solution points. At each 
solution point, in shared block $t_y$, the marking is read from the shared space, and slope limiting is applied if 
this marking is greater than zero. 

\section{Results}

Two test cases are presented for both a smooth and discontinuous problem. For all methods, a three 
state Runge-Kutta \cite{rk} time stepping scheme was applied and the interface fluxes 
were evaluated using the Rusanov \cite{rusanov} Riemann solver. The time step for each method 
was computed using the $CFL$ condition as
\begin{equation}
 \Delta t \le \frac{CFL \Delta x}{|u| + c}.
\end{equation}
The $CFL$ number for high-order methods is known to be quite restrictive in comparison to FV. To ensure a fair comparison 
with FV, the following convention is used: At the end of a simulation, the error is recorded. A new simulation 
is completed at a value of $0.5*CFL$ of the previous. Again, the error is recorded. If the percent error between 
these two errors is less than $0.1 \%$, the $CFL$ is termed the maximum $CFL$. Errors were completed by comparing the 
averaged solution with the averaged exact solution. For $P^2$ FV, the error was computed by reconstructing the 
solution along element faces, and then using a quadrature rule to compute an averaged solution \cite{ho}.
Finally, since FV has one 
solution state per element, while high-order methods have mutliple, the total number of degrees of freedom $NDoF$ 
between the methods was held constant. Only quadrilateral elements are considered in this work.
For one element and a $P^2$ reconstruction for a high-order method, 
$NDoF = 9$. To match this, the FV method must have 9 elements. 

A single Tesla K20c GPU card was used for all simulations. The code was 
compiled under compute architecture 3.5 using the CUDA toolkit version 6.0. In addition, the \verb|-O3| compiler 
optimization was used as well as the CUDA 64-bit libraries. Double precision is used for all computations.
The computational time was nondimensionalized by taubench \cite{tau} using 
the following taubench condition: \newline \verb|./Taubench -n 250000 -s 10|. On the GPU workstation used in our simulations, 
taubench gave a value of 8.274. This produces what is known as a work unit, 
as suggested by the 1st International Workshop on High-Order Methods \cite{ho} when comparing timings from numerical methods.
A work unit is a nondimensionalized unit computed by dividing the computational time it takes to complete a simulation by 
the taubench result.

\subsection{Smooth Problem}

A vortex propagation case is used as the smooth problem in this paper. The flow of the vortex is characterized in Ref.
\cite{shu}. A mean flow is specified $(\rho, u, v, p) = (1, 1, 0, 1)$ with fluctuation in the velocity, temperature ($T$), 
and entropy ($S$),
\begin{align*}
 (\delta u, \delta v) &= \frac{\epsilon}{2\pi} e^{0.5(1-r^2)} \left ( -y, x\right), \\
 \delta T &= -\frac{(\gamma-1)\epsilon^2}{8\gamma \pi^2}e^{1-r^2}, \\
 \delta S &= 0.
\end{align*}
Here, $r^2 = x^2 + y^2$ and the vortex has strength $\epsilon = 5$. An exact solution exists and can be found 
using $x_e = x - ut$ and $y_e = y - vt$, where $t$ is the final time. The solution evolves until time 
$t=1$ and the $L_2$ error norm of $\rho$ is computed. The domain is taken as $[-5, 5] \times [-5, 5]$ and periodic 
conditions are imposed on the boundaries. Discretizations from $20 \times 20$ to $100 \times 100$ quadrilateral 
elements are used in the simulations. 
Table 1 shows the maximum $CFL$ chosen for 
the runs, which allowed a less than $0.1\%$ error change when the $CFL$ was decreased by $1/2$. 
\begin{figure}[t]
\centering
\mbox{\subfigure[]{\includegraphics[width=2.4in]{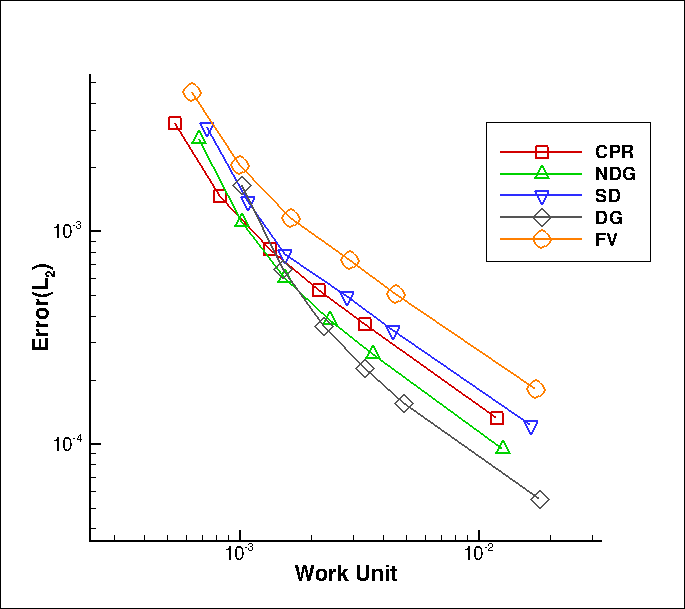}}\quad
\subfigure[]{\includegraphics[width=2.4in]{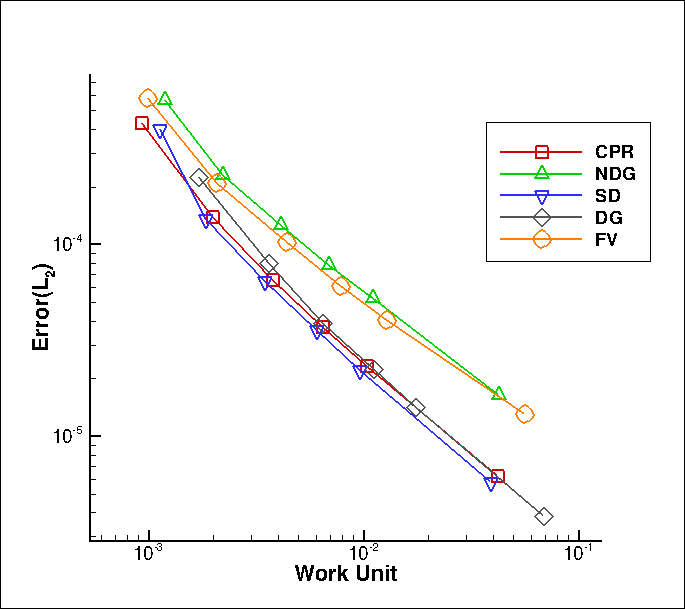} }} \\
\subfigure[]{\includegraphics[width=2.4in]{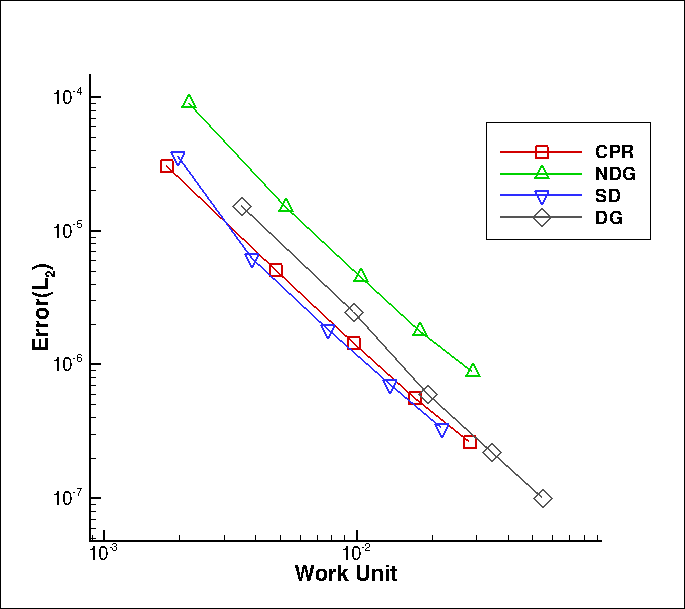} }
\caption{$L_2$ density errors using (a) $P^1$, (b) $P^2$, and (c) $P^3$ reconstructions versus work unit}
\label{wu_vor}
\end{figure}

\begin{table}\footnotesize
    \caption{Maximum $CFL$ - Smooth problem} \vspace{0.1in}
   \label{tab:1}
        \centering 
   \begin{tabular}{|c | c | c | c |  c | c |} 
      \hline 
      $P^1$|$DoFs$ & CPR & NDG & SD & DG & FV \\
       \hline 
       1600   &  0.24      & 0.24      &  0.3  & 0.24      & 0.4  \\
      3600   &  0.24      & 0.24      &  0.3  & 0.24      & 0.4  \\
      6400   &  0.24      & 0.24      &  0.3  & 0.24      & 0.38  \\
      10000   &  0.24      & 0.24      &  0.3  & 0.24      & 0.38  \\
      14400   &  0.24      & 0.24      &  0.3  & 0.24      & 0.37  \\
      \hline
       $P^2$|$DoFs$ & CPR & NDG & SD & DG & FV \\
      \hline 
      3600   &  0.14      & 0.14      &  0.2  & 0.14      & 0.4  \\
      8100   &  0.13      & 0.13      &  0.2  & 0.13      & 0.4  \\
      14400   &  0.13      & 0.13      &  0.2  & 0.13      & 0.38  \\
      22500   &  0.13      & 0.13      &  0.2  & 0.13      & 0.37  \\
      32400   &  0.13      & 0.13      &  0.2  & 0.13      & 0.37  \\
      \hline
   \end{tabular}
\end{table}

\begin{figure}[t]
\centering
\mbox{\subfigure[]{\includegraphics[width=2.4in]{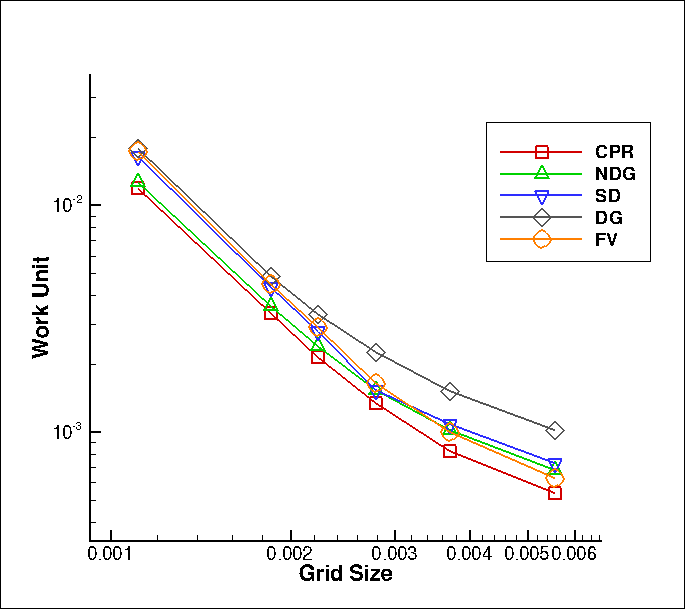}}\quad
\subfigure[]{\includegraphics[width=2.4in]{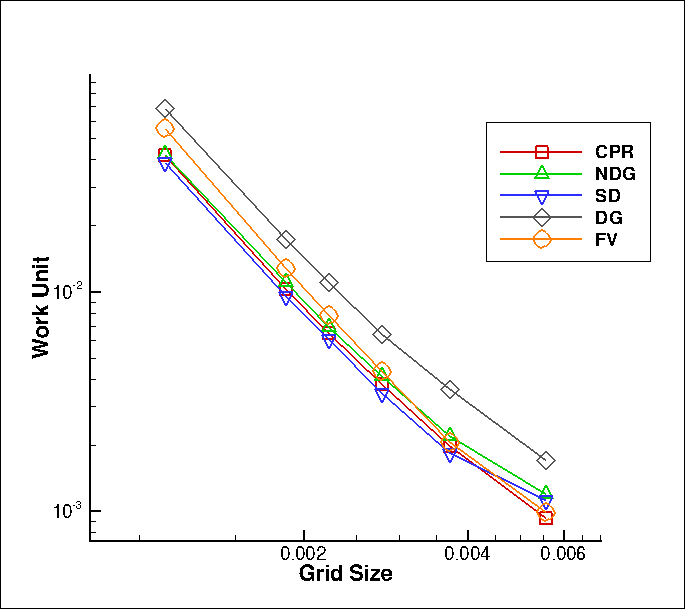} }} \\
\subfigure[]{\includegraphics[width=2.4in]{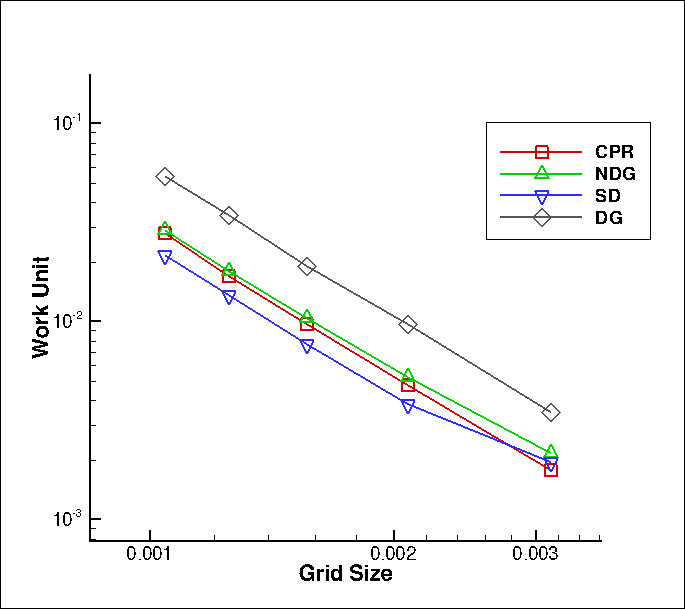} }
\caption{Total work unit to finish simulations for (a) $P^1$, (b) $P^2$, and (c) $P^3$}
\label{time_vor}
\end{figure}

The solution errors versus work units for $P^1$, $P^2$, and $P^3$ reconstructions are shown in Fig. \ref{wu_vor}.
The overall trend of increasing accuracy with increasing work unit is observed for all methods.
It is observed that the high-order methods obtain smaller error thresholds than the FV method for a given work unit (exception 
for $P^2$ NDG, which has larger errors associated). For $P^1$ reconstructions shown in Fig. \ref{wu_vor} (a), DG clearly outperforms 
other methods, but as the order is increased, Fig. \ref{wu_vor} (b) shows CPR and SD both achieve comparable errors with DG for a given 
work unit. In the case of CPR, the schemes compact nature is attributed to this, 
where the operations to compute the flux derivative are contained in one GPU kernel.
For SD, Table \ref{tab:1} showis that the SD method can take larger time-steps than the other high-order methods, as the CFL is not as restrictive.
Similar plots are observed in the comparative study done by Yu \textit{et al.} \cite{yu2}, where the error is compared with work units for several high-order methods on CPUs.
They also observe that CPR has a considerably lower error given a work unit for $P^2$ reconstruction than other methods, while NDG is significantly higher. 
In the results presented here, SD is comparable to CPR because the maximum allowable time-step for a given temporal error is used.
This is different than the approach in Ref. \cite{yu2}, where a constant time-step is implemented.
A fourth order reconstruction is also completed, and shown in Fig. \ref{wu_vor} (c). It illustrates that the SD
and CPR methods both obtain the lowest errors for a given work unit for a $P^3$ reconstruction.

Figure \ref{time_vor} shows the work unit needed to complete a simulation on a given mesh for $P^1$, $P^2$, and $P^3$ 
reconstructions. The obvious trend of the
work unit increasing for finer meshes is observed for both orders of accuracy.
To complete a full simulation, the FV and CPR methods are the fastest on coarse meshes (Fig. \ref{time_vor} (a) and (b)). 
Small computational domains do not take advantage of the GPU architecture with the optimizations and different memory 
types discussed in this paper. As the domain is refined and the order is, the high-order schemes can produce solutions faster than FV. 
The data illustrates that on fine meshes with high-order reconstruction, the high-order CPR, NDG, and SD methods run faster
than the FV method
as the degrees of freedom are increased. 

Further increasing the order to $P^3$, Fig. \ref{time_vor} (c), shows the CPR and NDG 
converge to the same work unit for a given simulation. The SD method, however, is able to complete solutions faster than 
any other high-order method for $P^3$ reconstruction. The solution errors are recorded for the high-order methods and are shown in 
Tables \ref{tab:6} and \ref{tab:7}. For $P^1$ reconstructions, the DG method produces the lowest $L_2$ errors and all methods slopes 
decay at a rate equivalent to the order of accuracy. A similar trend is observed for $P^2$ errors. Note that the NDG errors are significantly higher,
which is due to aliasing issues with the method. 

  \begin{table}\footnotesize
  \begin{center}
  \caption{High-order error values for smooth problem ($P^1$ reconstruction)} \vspace{0.1in}
   \label{tab:6}
  \begin{tabular}{|c | c | c | c |  c |}
    \hline
       Method & CPR Error & CPR Slope & NDG Error & NDG Slope  \\
      \hline 
      20$\times$20   &  3.21E-003 & -     &  2.70E-003  & -        \\
      30$\times$30   &  1.46E-003 & 1.94 &  1.10E-003 & 2.22   \\
      40$\times$40   &  8.25E-004 & 1.98 &  6.04E-004 & 2.07   \\
      50$\times$50   &  5.29E-004 & 1.99 &  3.83E-004 & 2.04   \\
      60$\times$60   &  3.67E-004 & 2.00 &  2.65E-004 & 2.02   \\
      \hline
      Method & SD Error & SD Slope & DG Error & DG Slope  \\
      \hline 
      20$\times$20   &  3.07E-003 & -    &  1.65E-003  & -        \\
      30$\times$30   &  1.38E-003 & 1.96 &  6.63E-004 & 2.24   \\
      40$\times$40   &  7.78E-004 & 2.00 &  3.59E-004 & 2.13   \\
      50$\times$50   &  4.96E-004 & 2.02 &  2.26E-004 & 2.08   \\
      60$\times$60   &  3.44E-004 & 2.02 &  1.55E-004 & 2.06   \\
    \hline
  \end{tabular}
    \end{center}
\end{table}

  \begin{table}\footnotesize
  \begin{center}
    \caption{High-order error values for smooth problem ($P^2$ reconstruction)} \vspace{0.1in}
   \label{tab:7}
   \begin{tabular}{|c | c | c | c |  c |}
   \hline
       Method & CPR Error & CPR Slope & NDG Error & NDG Slope  \\
      \hline 
      20$\times$20   &  4.30E-004 & -     &  5.63E-004  & -        \\
      30$\times$30   &  1.38E-004 & 2.81 &  2.30E-004 & 2.20   \\
      40$\times$40   &  6.51E-005 & 2.60 &  1.26E-004 & 2.10   \\
      50$\times$50   &  3.69E-005 & 2.55 &  7.81E-005 & 2.14   \\
      60$\times$60   &  2.32E-005 & 2.55 &  5.23E-005 & 2.19   \\
      \hline
      Method & SD Error & SD Slope & DG Error & DG Slope  \\
      \hline 
      20$\times$20   &  4.00E-004 & -    &  2.24E-004  & -        \\
      30$\times$30   &  1.36E-004 & 2.66 &  7.95E-005 & 2.55   \\
      40$\times$40   &  6.40E-005 & 2.62 &  3.90E-005 & 2.48   \\
      50$\times$50   &  3.57E-005 & 2.62 &  2.24E-005 & 2.48   \\
      60$\times$60   &  2.21E-005 & 2.62 &  1.42E-005 & 2.50   \\
    \hline
  \end{tabular}
    \end{center}
\end{table}

\subsection{Discontinuous Problem}

\begin{figure}[t]
\centering
\mbox{\subfigure[]{\includegraphics[width=2.4in]{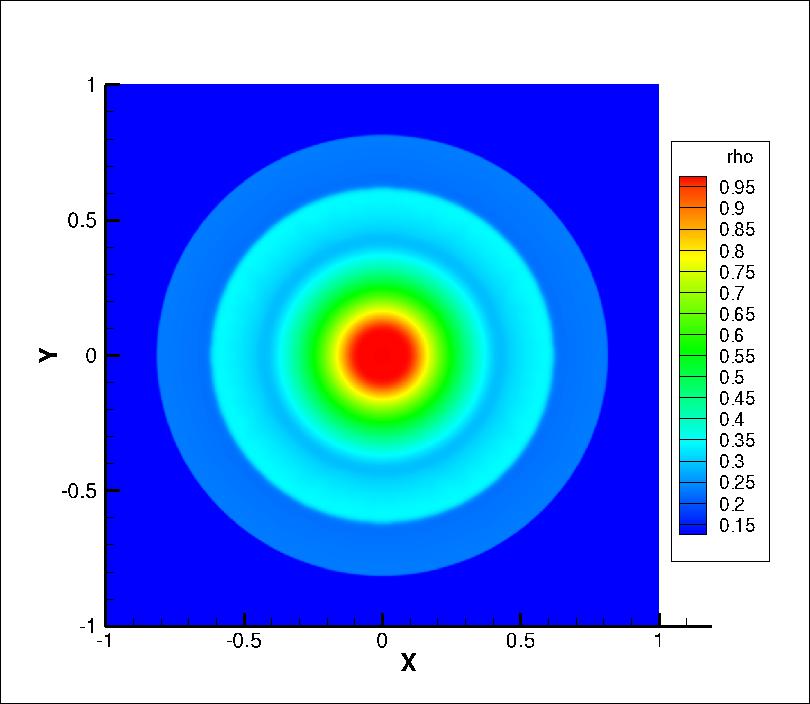}}\quad
\subfigure[]{\includegraphics[width=2.4in]{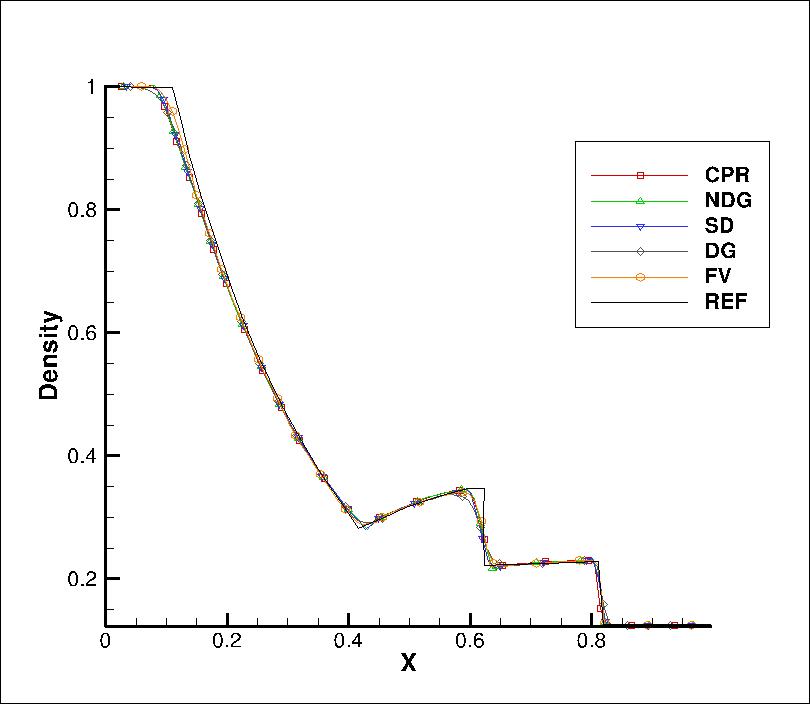} }} \\
\caption{Discontinuous test case results (a) Density contours (b) $P^1$ solution comparison}
\label{blast}
\end{figure}

The next case is a radially expanding shock tube from Toro \cite{toro}. A domain of size $[-1,1] \times [-1,1]$
initializes density 
and pressure $(\rho, p)$ of $1.0$ inside a radius of $0.4$. Outside the radius, $\rho = 0.125$ and $p = 0.1$. 
There is no velocity component at the initial time. Rather than using solution errors to check if the CFL is 
small enough, the residual error is used. 
\begin{table}\footnotesize
	\fontsize{10}{10}\selectfont
    \caption{Maximum CFL - Discontinuous problem} \vspace{0.1in}
   \label{tab:3}
        \centering 
   \begin{tabular}{|c | c | c | c |  c | c |} 
      \hline 
       $P^1$|$DoFs$ & CPR & NDG & SD & DG & FV \\
      \hline 
      160k   &  0.2      & 0.2      &  0.3  & 0.22      & 0.58  \\
      640k   &  0.2      & 0.2      &  0.27  & 0.2      & 0.58  \\
      \hline
      $P^2$|$DoFs$   &  CPR & NDG & SD & DG & FV \\
\hline
      360k   &  0.1      & 0.1      &  0.18  & 0.08      & 0.54  \\
      1440k   &  0.1      & 0.1      &  0.18  & 0.08      & 0.54  \\
      \hline
   \end{tabular}
\end{table}
%
The solution is ran until a final time of $t=0.25$, where the density is compared along the centerline, $y=0$. For the reference solution, 
the data was taken from the text Riemann Solvers and Numerical Methods for Fluid Dynamics \cite{toro} (digitized for use here).
As illustrated in Fig. \ref{blast} (b), all methods have good agreement with the reference solution. 

\begin{figure}[t]
\centering
\mbox{\subfigure[]{\includegraphics[width=2.4in]{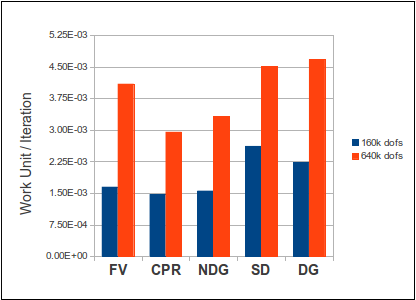}}\quad
\subfigure[]{\includegraphics[width=2.4in]{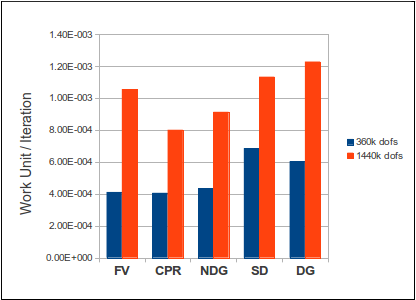} }} \\
\caption{Computational work per iteration for (a) $P^1$ and (b) $P^2$ reconstruction}
\label{tpip1}
\end{figure}
%

Fig. \ref{tpip1} illustrates the work unit needed per iteration for each method
for both $P^1$ and $P^2$ reconstructions.
As the computational domain is increased, the benefit of using some high-order methods becomes apparent. 
For the $P^2$ reconstruction in Fig. \ref{tpip1} (b) with $1440$k degrees of freedom, CPR is $27\%$ faster per iteration than FV, 
while DG is $14\%$ slower than FV.

\begin{figure}[t]
\centering
\mbox{\subfigure[]{\includegraphics[width=2.4in]{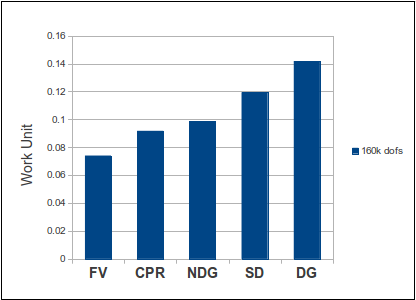}}\quad
\subfigure[]{\includegraphics[width=2.4in]{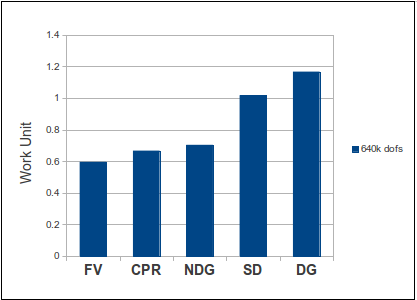} }} \\
\caption{Total work for $P^1$ reconstruction (a) 160,000 degrees of freedom and (b) 640,000 degrees of freedom}
\label{f1}
\end{figure}

\begin{figure}[t]
\centering
\mbox{\subfigure[]{\includegraphics[width=2.4in]{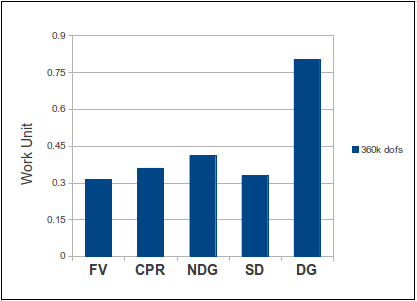}}\quad
\subfigure[]{\includegraphics[width=2.4in]{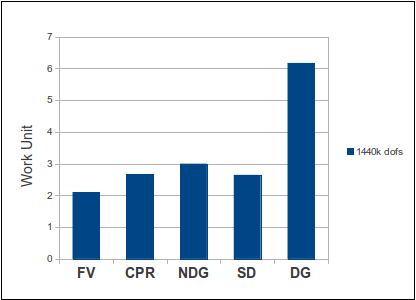} }} \\
\caption{Total work for $P^2$ reconstruction (a) 360,000 degrees of freedom and (b) 1,440,000 degrees of freedom}
\label{f2}
\end{figure}


The total computational work is shown in Fig. \ref{f1} and \ref{f2}.
For $P^1$ reconstructions, the CPR method is the fastest of the high-order methods, while SD and 
DG take the most time. Once the reconstruction is increased to $P^2$, a similar trend is observed, however 
the total work for SD and CPR is nearly identical. This is due to the SD schemes ability to take larger time steps than the CPR 
approach. However, the time step restrictions on high-order methods are harsher than FV, which enables FV to arrive at the final solution time 
$25\%$ faster than the CPR/SD schemes for a $P^2$ reconstruction with $1440$k degrees of freedom. This is also the case 
because of the extra sweeps the high-order methods need to take for discontinuous problems, one for evaluating the average and 
another to apply the limiting procedure.
%

\section{Conclusions}

The presented work compares multiple numerical methods implemented on GPUs using CUDA computing. The algorithms for 
each method were presented, and used to test smooth and discontinuous problems. The maximum allowable time step
from the CFL condition was used for each problem, with the number of degrees of freedom held constant across the 
methods. For the smooth problem, the high-order methods obtained an error threshold at a lower work unit 
than the FV method. Additionally, the CPR, NDG, and SD methods are capable of arriving at solutions faster than the 
FV approach as the computational domain is increased. For discontinuous problems, the FV method does
produce solutions $25\%$ faster than the fastest high-order methods, but solution profiles between the methods are similar. The computational work per step shows 
the CPR and NDG methods are most efficient, but time-step restrictions cause slower solution generation when compared to FV.

This two-dimensional approach may provide a foundation for comparisons with three-dimensional methods.
The extension is necessary because bottlenecks shift when going from two-dimensional to three-dimensional problems.
Additionally, different grids and unstructured mesh cases should be considered.

\section*{Acknowledgements} 
This research has been supported   by a NIAC (NASA Innovative Advanced Concepts) Phase 2 study entitled
``An Innovative Solution to NASA's Asteroid Impact Threat Mitigation Grand Challenge and Its Flight Validation Mission Design."
Additional support has been given by the Vance Coffman Chair Fund. Approved for unlimited release: LA-UR-17-27957.


\bibliographystyle{h-physrev}
\bibliography{references}{}

\begin{thebibliography}{10}

\bibitem{ho}
Z.~J. Wang {\em et~al.},
\newblock Int J Numer Methods Fluids {\bf 72}, 811 (2013).

\bibitem{zjsv}
Z.~J. Wang and H.~Gao,
\newblock J Comput Phys {\bf 178}, 210 (2002).

\bibitem{castro}
M.~Castro, S.~Ortega, M.~de~la Asunci\'{o}n, J.~M. Mantas, and J.~M. Gallardo,
\newblock High Performance Computing {\bf 339}, 165 (2011).

\bibitem{oben}
K.~Obenschain, K.~Corrigan, and G.~Patnaik,
\newblock AIAA  (2011).

\bibitem{bass}
F.~Bassi and S.~Rebay,
\newblock J Comput Phys {\bf 138}, 251 (1997).

\bibitem{bau}
C.~E. Baumann and T.~J. Oden,
\newblock J Numer Meth Fluids {\bf 31}, 79 (1999).

\bibitem{cock}
B.~Cockburn and C.~W. Shu,
\newblock J Comput Phys {\bf 141}, 199 (1998).

\bibitem{cock0}
B.~Cockburn and C.~W. Shu,
\newblock Math Comput {\bf 52}, 411 (1989).

\bibitem{cock00}
B.~Cockburn, S.~Lin, and C.~W. Shu,
\newblock J Comput Phys {\bf 84}, 90 (1989).

\bibitem{reed}
W.~H.~Reed and T.~R.~Hill,
\newblock Los Alamos Scientific Laboratory Report, 1973 (unpublished).

\bibitem{Hes}
J.~S. Hesthaven and T.~Warburton,
\newblock {\em Nodal {D}iscontinuous {G}alerkin {M}ethods: {A}lgorithms,
  {A}nalysis, and {A}pplications.} (Springer-Verlag, New York, 2008).

\bibitem{klock}
A.~Kl{\"o}ckner, T.~Warburton, J.~Bridge, and J.~S. Hesthaven,
\newblock J Comput Phys {\bf 228}, 7863 (2009).

\bibitem{huynh}
H.~T. Huynh,
\newblock AIAA  (2007).

\bibitem{wangbook}
Z.~J. Wang,
\newblock {\em Adaptive {H}igh-{O}rder {M}ethods in {C}omputational {F}luid {D}ynamics} (2011).

\bibitem{gao}
Z.~J. Wang and H.~Gao,
\newblock J Comput Phys {\bf 228}, 8161 (2009).

\bibitem{yu}
M.~L. Yu and Z.~J. Wang,
\newblock J Sci Comput {\bf 54 (1)}, 227 (2013).

\bibitem{hoff1}
M.~Hoffmann, C.-D. Munz, and Z.~J. Wang,
\newblock ICCFD  (2012).

\bibitem{zimm}
B.~J. Zimmerman and Z.~J. Wang,
\newblock Comput Fluids {\bf 101}, 263 (2014).

\bibitem{liu06}
Y.~Liu, M.~Vinokur, and Z.~J. Wang,
\newblock J Comput Phys {\bf 216}, 780 (2006).

\bibitem{may}
G.~May and A.~Jameson,
\newblock AIAA  (2006).

\bibitem{sun}
Y.~Sun and Z.~J. Wang,
\newblock J Comput Phys {\bf 2}, 301 (2007).

\bibitem{zimm2}
B.~J. Zimmerman, Z.~J. Wang, and M.~Visbal,
\newblock AIAA  (2013).

\bibitem{yu2}
M.~L. Yu, Z.~J. Wang, and Y.~Liu,
\newblock J Comput Phys {\bf 259}, 75 (2014).

\bibitem{muscl1}
B.~Van~Leer,
\newblock J Comput Phys {\bf 14}, 361 (1974).

\bibitem{muscl}
B.~Van~Leer,
\newblock J Comput Phys {\bf 32}, 101 (1979).

\bibitem{harten}
A.~Harten, B.~Enquist, S.~Osher, and S.~R. Chagravarthy,
\newblock J Comput Phys {\bf 71}, 231 (1987).

\bibitem{cock9}
B.~Cockburn and C.~W. Shu,
\newblock J Sci Comput {\bf 16}, 173 (2001).

\bibitem{kub}
E.~Kubatko, C.~Dawson, and J.~Westerink,
\newblock J Comput Phys {\bf 227}, 9697 (2008).

\bibitem{jac}

\newblock O.~C. Zienkiewicz and R.~C. Tayler {\em The {F}inite {E}lement
  {M}ethod the {B}asics} Vol.~1 (2000).

\bibitem{tan}

\newblock J.~C. Tannehill, A.~A. Anderson, and R.~H. Pletcher {\em Computational
  {F}luid {M}echanics and {H}eat {T}ransfer} Vol.~2 (1997).

\bibitem{minmod}
P.~L. Roe,
\newblock Rev in Fluid Mech {\bf 18}, 337 (1986).

\bibitem{nvidia}
\newblock {\em {NVIDIA} {CUDA} {C} {P}rogramming {G}uide} Vol. 5.0 (2012).

\bibitem{zimmThesis}
B.~J. Zimmerman,
\newblock The efficient implementation of correction procedure via
  reconstruction with {GPU} computing,
\newblock Master's thesis, Iowa State University, 2013.

\bibitem{rk}
C.~W. Shu,
\newblock SIAM J Sci Stat Comput {\bf 9}, 1073 (1988).

\bibitem{rusanov}
V.~V. Rusanov,
\newblock J Comput Math Phys {\bf USSR(1)}, 267 (1961).

\bibitem{tau}
Taubench.

\bibitem{shu}
C.~W. Shu,
\newblock {\em Essentially non-oscillatory and weighted essentially
  non-oscillatory schemes for hyperbolic conservation laws, in: {A}dvanced
  {N}umerical {A}pproximation of {N}onlinear {H}yperbolic {E}quations}
  (Springer-Verlag, Berlin/New York, 1998).

\bibitem{toro}
E.~F. Toro,
\newblock {\em Riemann {S}olvers and {N}umerical {M}ethods for {F}luid
  {D}ynamics: {A} {P}ractical {I}ntroduction} (Springer, London/New York,
  2009).

\end{thebibliography}
\end{document}